\documentclass[useAMS,usenatbib]{mn2e}
\usepackage{aas_macros}
\usepackage{graphicx}
\usepackage{color}
\newcommand{\gadg}{\textsc{gadget }}
\newcommand{\gadgt}{\textsc{gadget-$3$}}
\newcommand{\subf}{\textsc{subfind }}
\newcommand{\fof}{\textsc{fof }}

\title[Adaptive softening in GADGET]{Adaptive gravitational softening in GADGET}
\author[F.Iannuzzi and K. Dolag]{Francesca Iannuzzi$^{1}$\thanks{E-mail:
iannuzzi@mpa-garching.mpg.de} and Klaus Dolag$^{1,2}$\\
$^{1}$Max-Planck Institut f\"{u}r Astrophysik, Karl-Schwarzschild Str. 1, D-85741 Garching, Germany\\\
$^{2}$University Observatory Munich, Scheinerstr. 1, 81679 M\"{u}nchen, Germany}
\begin{document}

\date{Accepted 2011 July 13. Received 2011 July 13; in original form 2011 April 18}

\pagerange{\pageref{firstpage}--\pageref{lastpage}} \pubyear{2011}

\maketitle

\label{firstpage}

\begin{abstract}
Cosmological simulations of structure formation follow the
collisionless evolution of dark matter starting from a nearly
homogeneous field at early times down to the highly clustered
configuration at redshift zero. The density field is sampled by a
number of particles in number infinitely smaller than those believed
to be its actual components and this limits the mass and spatial
scales over which we can trust the results of a simulation. Softening of
the gravitational force is introduced in collisionless simulations to
limit the importance of close encounters between these particles. The scale of
softening is generally fixed and chosen as a compromise between the
need for high spatial resolution and the need to limit the particle
noise. In the scenario of cosmological simulations, where the density
field evolves to a highly inhomogeneous state, this compromise results
in an appropriate choice only for a certain class of objects, the
others being subject to either a biased or a noisy dynamical
description. We have implemented adaptive gravitational softening
lengths in the cosmological simulation code \gadgt; the formalism
allows the softening scale to vary in space and time according to the
density of the environment, at the price of modifying the equation of
motion for the particles in order to be consistent with the new dependencies
introduced in the system's Lagrangian. We have applied the technique to a number of
test cases and to a set of cosmological simulations of structure
formation. We conclude that
the use of adaptive softening enhances the clustering of particles
at small scales, a result visible in the amplitude of the
correlation function and in the inner profile of massive objects,
thereby anticipating the results expected from much higher 
resolution simulations.
\end{abstract}


\section{Introduction}
In collisionless N-body simulations the matter field is represented by a number of point-particles in number considerably reduced with respect to the actual building blocks of the system. These particles have no physical counterpart and should be simply regarded as Monte-Carlo samplings of the probability-density distribution in position and velocity. When during the evolution two of these particles are found at small spatial separations, the gravitational attraction between them can become arbitrarily large and following the encounter properly can turn out to be prohibitively expensive from the computational point of view; plus, this collisional behaviour is an artifact of the granularity of the mass distribution in the simulation and would not manifest in a intrinsically-smooth, collisionless system. A way around this problem, namely the discreetness of the particle representation, is obtained by smoothing the density distribution from a collection of spikes to a continuous field; to this purpose, every particle is assigned a finite volume over which its mass is spread. This translates into the gravitational interaction between the computational particles being ``softened'' at small separations, i.e. being attenuated and prevented to diverge. In this way close encounters are impeded and the simulation can proceed at a regular pace. The gravitational smoothing does not considerably affect the artificial two-body relaxation though, as this is driven by both close and distant encounters and the former cannot be avoided by softening techniques \citep{chandra42, spitzer71, hernquist_barnes90, theis98, dehnen01, diemand04}. Suppressing relaxation is primarily a condition on the number of sampling particles and depends only weakly on the value of the softening \citep{power2003}. \\
The details of this smoothing procedure can be reduced to the choice of a softening kernel $W(r,h)$, which determines the functional form of the modified density profile of the particle, and of a softening length $h$, which controls  the spatial extent within which the modification applies. In the classic case of ``Plummer'' softening, the gravitational potential of each particle is that of a Plummer sphere whose scale length is given by the value of the softening $h$; this results in the following form for the gravitational force $\bmath{F}(\bmath{r})$ between two particles of masses $m_i$ and $m_j$ separated by a distance $\bmath{r}$:
\begin{equation}\label{eq:plummer_softening}
\bmath{F}(\bmath{r}) = -G \frac{m_i m_j}{(r^2 +h^2)^{3/2}}\bmath{r}.
\end{equation}
An evident shortcoming of this choice is that the gravitational force is modified at all separations; several works (e.g. \citealt{dyerip93}; \citealt{dehnen01}) have shown that a better choice is to use kernels with a compact support, meaning that the interaction recovers its Newtonian, original form at separations grater than the softening length. The most commonly used one is the cubic spline of \cite{ml85}:
\begin{equation}\label{eq:kernel_ml85}
W(r,h) = \frac{8}{(\pi h^3)}\left\{
\begin{array}{lcclc}
1-6q^2+6q^3, &\quad& 0   &\leq q <& 0.5\\
2(1-q)^3,    &\quad& 0.5 &\leq q <& 1\\
0,           &\quad& 1   &\leq q
\end{array}\right.
\end{equation}
where $q=r/h$, $r$ being the distance from the centre of the kernel, i.e. from the particle's position. The particles assume the density profile given by Eq.~\ref{eq:kernel_ml85} and the gravitational potential and force fields are modified accordingly.\\
It is evident that the introduction of softening leads to an unphysical modification of the gravitational interaction at interparticle separation below $h$; the inverse square law is replaced by a gentler interaction, whose strength approaches zero in the limit $r \rightarrow 0$. This results in a systematic misrepresentation of the force at small scales, an effect commonly referred to as \textsl{bias} \citep{merritt96,dehnen01}. Ideally one would like to limit this effect to the smallest possible scales by reducing $h$ accordingly; at the same time the softening lengths cannot be made arbitrarily small without running into the discreetness problem previously addressed. It is then clear how the choice of $h$ assumes a critical importance for the reliability of the force computation and indeed a number of studies \citep{merritt96,romeo98,athanassoula00,dehnen01} have been devoted to the subject.\\
In general there is an overall agreement that no such a thing as an
``optimal'' softening length exists when one has to deal with highly
inhomogeneous systems, as too big a value would degrade the
description of over-dense regions whilst too small a value would
enhance collisionality in under-dense environments. Generally, the
gravitational softening scale is set to a fixed value, indeed decided as a compromise between the desired spatial resolution and the need to moderate the noise arising from particle-particle interactions. As a matter of fact, in most cases this means fine-tuning the softening parameter to follow properly the dynamics of a certain class of objects \-- the densest \-- at a certain time; in a cosmological simulations, where the initially uniform matter density field evolves to form highly differentiated structures, this implies following poorly not only the moderately-dense regions, but also the progenitors of the ``target'', densest objects.\\
Ideally it would be desirable to vary the softening scale in space and time according to the density evolution of the different environments: this would remove the aforementioned trade-off between spatial resolution and particle noise, thus allowing to increase the former and reduce the latter depending on the local features of the density field. The problem with a varying softening length though, as will be shown in Section~\ref{sec:formalism}, is that it introduces an additional dependence on the spatial position $r$ in the definition of the gravitational potential; if this is not accounted for in the derivation of the equation of motion, the energetics of the system can manifest unwanted behaviours. In this context, \cite{PM07} (hereafter PM07) have proposed a formalism to adapt the gravitational softening lengths in a simulation while retaining conservation of both momentum and energy, which, as just mentioned, would be lost when adopting an adaptive scheme. In their method the softening lengths are computed in the same fashion as for the smoothing length in smoothed particle hydrodynamics (SPH - \citealt{gm77}; \citealt{lucy77}), where $h$ varies with the local density $\rho$ according to $h\propto \rho^{-1/3}$. The adoption of this scheme involves a number of modifications in the overall structure of a typical cosmological simulation code, most noticeably the introduction of an additional term in the equations of motion. The technique has already been employed in  simulation codes like the TreePM code of \cite{BK09} (hereafter BK09) and the TreeSPH \textsc{EvoL} code of \cite{merlin}. BK09 apply the method to collisionless simulations of structure formation and find an increase in clustering in over-dense regions accompanied by a somewhat loss in resolution in the under-dense ones; \cite{merlin} performed standard hydrodynamical tests in order to test the overall performance of their code and found good energy-conservation properties along with an improved behaviour in tests like the isothermal collapse of \cite{bate97}.
Although we are here concerned about particle-based codes, where the gravitational interaction is computed at the particle's location, we note that grid-based codes employing adaptive mesh refinement \citep{bryan97,truelove98,abel00,knebe01,kravtsov02, teyssier02,oshea04,quilis04} are intrinsically spatially adaptive; these codes perform a recursive refinement of an initial regular grid, thereby increasing their resolution in regions of interest. For similar reasons as those outlined above, conservation of energy is not strictly guaranteed in these codes, as it would not be when using adaptive gravitational softening without changing the equation of motion accordingly.\\
In this paper we discuss the implementation of the adaptive softening
length formalism by PM07 in the cosmological simulation code \gadg
(\citealt*{springel01b}; \citealt{springel05}) and the main effects
that it has on dark matter simulations of the large-scale structure of
the Universe. Our results qualitatively confirm those of BK09 in
over-dense regions, but show no drawbacks in under-dense ones; in particular, we do not register a significant loss of low-mass structures and we considerably improve the results in term of particle clustering when comparing to standard simulations employing fixed softening.\\
In Sections~\ref{sec:formalism} and~\ref{sec:gadget} we briefly describe
the technique and its implementation in \gadg; in Section~\ref{sec:tests} we show two of the tests performed to check
the implementation and its result in controlled scenarios for which we know the solution, whereas in Sections~\ref{sec:cosmo} and \ref{sec:mmII} we apply the technique to cosmological simulations of structure formation. We summarise the results and conclude in Section~\ref{sec:concl}.

\section[]{The formalism}
\label{sec:formalism}
In this section we will briefly report the gist of the algorithm. We refer the reader to the pivotal work PM07 for a thorough treatment of the problem and for a detailed derivation of the new equation of motion.\\
The Lagrangian for a system of $N$ particles interacting only through gravity reads:
\begin{equation}\label{eq:lagrange}
L = \sum_{i=0}^N m_i\left(\frac{1}{2}\bmath{v}_i^2 - \Phi_i\right),
\end{equation}
where
\begin{equation}\label{eq:potential}
\Phi_i(\bmath{r}_i)= -G \sum_{j=0}^{N} m_j \phi \left(\left|\bmath{r}_i - \bmath{r}_j\right|, h\right).
\end{equation}
The expression for the potential $\Phi$ departs from its Newtonian definition due to the introduction of softening and hence of the kernel $\phi$ and the related length scale $h$. When this quantity has no further dependences and is kept fixed the resulting equation of motion will resemble the Newtonian original, but for the inheritance of this additional scale; its expression is derived by applying the Euler-Lagrange equations to (\ref{eq:lagrange}) and for particle $i$ it will read:
\begin{equation}
m_i\frac{d\mathbf{v}_i}{dt} = \sum_{j=0}^N \bmath{F}_{ij}(h),
\end{equation}
where 
\begin{equation}
\bmath{F}_{ij}(h)= -\nabla \Phi_{ij} =-G m_j \phi^\prime \left(\left|\bmath{r}_i - \bmath{r}_j\right|, h\right)\frac{\bmath{r}_i - \bmath{r}_j}{\left|\bmath{r}_i - \bmath{r}_j\right|}.
\end{equation}
Here $\phi^\prime = \partial \phi / \partial\left|\bmath{r}_i - \bmath{r}_j\right|$ is the force kernel. The expressions for both the potential and the force kernel in the cubic spline case can be found in BK09 (Appendix A) . The force $\mathbf{F}_{ij}(h)$ reduces to the Newtonian gravitational force for separations $\left|\bmath{r}_i - \bmath{r}_j\right| > h$ and to the kernel-dependent ``softened'' version otherwise.\\
When $h$ is allowed to vary from particle to particle according to the density of the environment a few complications arise. In order to maintain its translational invariance and hence ensure that the resulting dynamics would conserve linear momentum, the Lagrangian needs to be symmetrised. One way to do this is by averaging over the potentials evaluated with the softenings of the interacting pair of particles; the new Lagrangian would read:
\begin{equation}\label{eq:lagrange_symm}
L = \sum_{i=0}^N \frac{1}{2}m_i\bmath{v}_i^2 - \frac{G}{2}\sum_{i=0}^N
\sum_{j=0}^N   m_i m_j \left[ \frac{\phi_{ij}(h_i) + \phi_{ij}(h_j)}{2}\right],
\end{equation}
where $\phi_{ij}(h_i) \equiv \phi \left(\left|\bmath{r}_i - \bmath{r}_j\right|, h\right)$.
When deriving the equation of motion from (\ref{eq:lagrange_symm}),
the additional dependence on the position $r$ through $h$ results in an extra term
besides the classical inverse square law. The evolution of a particle $i$
subject only to the action of gravity and supplied with a variable
softening length will in fact obey:
\begin{eqnarray}\label{eq:eom}
\frac{d\bmath{v}_i}{dt} & = & -G \sum_{j=1}^{N} m_j\left[ \frac{\phi^\prime_{ij}(h_j) + \phi^\prime_{ij}(h_i)}{2}\right] \frac{\bmath{r}_i - \bmath{r}_j}{\left|\bmath{r}_i - \bmath{r}_j\right|} \nonumber \\
 & & -\frac{G}{2} \sum_{j=1}^{N} m_j\left[ \frac{\zeta_i}{\Omega_i} \frac{\partial W_{ij}(h_i)}{\partial\bmath{r}_i} + \frac{\zeta_j}{\Omega_j} \frac{\partial W_{ij}(h_j)}{\partial\bmath{r}_i}\right], 
\end{eqnarray}
where
\begin{equation}\label{eq:zeta}
\zeta_i \equiv \frac{\partial h_i}{\partial \rho_i}\sum_{k=0}^N m_k \frac{\partial \phi_{ik}(h_i)}{\partial h_i},
\end{equation}
\begin{equation}\label{eq:omega}
\Omega_i \equiv 1 - \frac{\partial h_i}{\partial \rho_i}\sum_{k=0}^N m_k \frac{\partial W_{ik}(h_i)}{\partial h_i}.
\end{equation}
The additional contribution, which we will refer to as the
``correction term'', is attractive in nature and will therefore act towards
increasing the resulting gravitational force. If this term were not
accounted for when using adaptive softening, the particles would be
evolved according to a law incoherent with the system's Lagrangian,
i.e. energy conservation would be lost.

\section[]{Implementation in Gadget}
\label{sec:gadget}
We have implemented the adaptive softening formalism on the latest version of the \gadg  code, namely \gadgt. \gadgt~computes gravitational forces via the TreePM method (\citealt{xu95}; \citealt*{bode00}; \citealt{bagla02}) and hydrodynamical interactions by means of SPH; it differs from the previous versions of the code in that it features a more flexible domain decomposition, something that made it suitable for simulations involving extreme levels of clustering, as the Millennium-II \citep{mII}.\\
Implementing adaptive softening required modifications of the Tree-algorithm and of the timestep criterion along with the introduction of the machinery to compute the softening lengths. Besides the obvious change in the expression for the gravitational acceleration, the opening criterion for the nodes has been modified in order to avoid the presence of smoothed particle-node interaction; in other words, the correction term can be non-zero only within a particle-particle interaction (more details in Sec. 3.2.4 of BK09). The timestep criterion employed by \gadg for collisionless particles reads:
\begin{equation}\label{eq:timecriterion}
\Delta t = \left( \frac{2\eta\epsilon}{a}\right)^{1/2}
\end{equation}
where $\eta$ is an accuracy parameter, $\epsilon$ is the Plummer equivalent softening ($h\simeq 2.8 \epsilon$, where $h$ is the support of the cubic spline kernel) and $a$ the acceleration of the particle. We kept the criterion itself unchanged and substituted $\epsilon$ with the individual value of the adaptive softening. \\
As to the computation of the softening lengths the method is identical to the one \gadg uses for setting the smoothing lengths in SPH calculations. Qualitatively, they represent the radius of the sphere centred on the particle and containing a specific number of companion particles, generally referred to as ``neighbours''; formally, this translates into the following relation:
\begin{equation}\label{eq:what.is.h}
\frac{4\pi}{3}h^3_i \sum_{j=0}^N W_{ij}(h_i) = N_{ngbs}
\end{equation}
where $N_{ngbs}$ stands for the number of neighbours and $N$ is the number of particles within distance $h_j$ from the target particle $j$. The above equation is solved iteratively with a Newton-Raphson method until the difference between the two sides falls below a certain tolerance threshold.\\
Adaptive softening lengths can be activated both when the code works in TreePM mode and when it uses the Tree-only algorithm. In the latter case the softening lengths are left varying without boundaries according to the local features of the particle distribution; in the former case we do instead allow for the presence of a minimum and maximum value for the softening lengths. As explained in Sec.~$3.2.1$ of BK09, the existence of a minimum value is not crucial and only prevents the simulation from becoming overly expensive in terms of computational time; conversely, the upper bound is introduced to ensure that the long-range force (the particle-mesh contribution) is negligible on the scales where softening is important, so that errors arising from the non-modification of the long range force are under control. These bounds are expressed in terms of the splitting scale $r_s$, the scale (generally of order the grid spacing) where the splitting of the potential in a long-range and short-range component is performed; choosing $h_{max} \simeq \frac{r_s}{2}$ results in the long-range contribution being below $1\%$ of the total force at scales where softening is important. Although we always impose a lower limit to the softening length when using the code in its TreePM mode, we did not find the presence of an upper limit to have dramatic consequences on the results, especially when using the adaptive formalism in its full, conservative version. 

\section[]{Tests}
\label{sec:tests}
In this section we present some of the tests performed in order to
check the correctness of the implementation and explore the general
effects of adaptive softening when simulating different physical scenarios. We will initially show the behaviour of the code in simulating simple systems, of well-known properties; the force profile of a Plummer and Hernquist models are investigated, and their temporal evolution in and out of equilibrium; the density profile of a polytrope and the behaviour of its total energy in time is also shown. Most of these examples are present already in PM07 and were specifically chosen to test our implementation.\\
In all the numerical simulations presented in this section we adopt units of mass $[M]=1$, length $[R]=1$ and $G=1$. As a result, the energy per unit mass is measured in units of $GM/R$ and time in $(GM/R^3)^{-1/2}$.  \\
\subsection{Evolution of a Plummer sphere}
The system considered here consists of a set of $N$ particles distributed according to a ``Plummer'' profile:
\begin{equation}
\rho(r) = \frac{3GMr_s^2}{4\pi(r_s^2 + r^2)^{5/2}}
\end{equation}
Here the total mass $M$ and the scale radius $r_s$ are set equal to unity.
The idea is to evaluate the resulting gravitational force profile and investigate its dependence on the choice of softening. Once this is accomplished we concentrate on the behaviour of the total energy as the system is let evolve in time.\\
This test is identical to that presented by PM07 and we refer to section 4.3 of their paper or alternatively to \citet{aarseth74} for details on the setup of the initial conditions.\\
The test was run using different number of particles and both Tree-only and TreePM algorithms for the evaluation of the gravitational force; the results behave as expected in the different cases and here we show only those obtained using the pure Tree method on $N=1000$ particles.\\
\begin{figure}
\includegraphics[width=84mm]{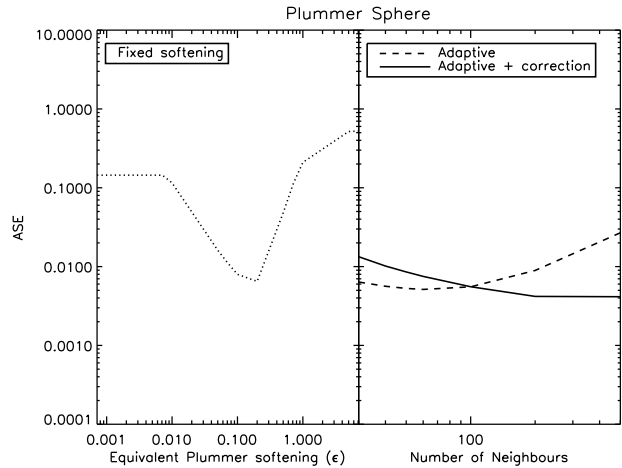}
 \caption{Average squared errors (ASE) of the force field generated by distributing $N=1000$ particles according to a Plummer profile. The left panel shows the behaviour of the ASE as a function of the choice of (fixed) softening; the right panel shows the results of adaptive softening varying the number of neighbours.}
  \label{ase}
\end{figure}
Fig.~\ref{ase} shows the averaged square errors (ASE) of the simulated force field corresponding to different choices of both fixed and adaptive gravitational softening. This quantity measures the deviation of the force experienced by particles at different radii from the analytical value, given by
\begin{equation}
f(r) = \frac{GMr}{(r_s^2 + r^2)^{3/2}} ,
\end{equation}
and it is defined as
\begin{equation}
\textrm{ASE} = \frac{1}{f_{max}^2 N} \sum_{i=1}^N \left|f_i -f_{exact}(\bmath{r_i})\right|^2,
\end{equation}
where $f_i$ is the force on particle $i$ and $f_{max}$ is the maximum value of the exact solution. For a discussion on the use of the ASE or related quantities to assess the error on a force profile we refer to PM07, \citet{merritt96} and \citet{dehnen01}.\\
The results show how less sensitive the representation of the force field is to the choice of $N_{ngbs}$ than to the choice of $\epsilon$; choosing from $30$ to $500$ neighbours does not shift the global ASE much away from the minimum value corresponding to the ``optimal'' choice of fixed softening ($\epsilon_{opt} \approx 0.2$, of the order the mean interparticle separation within the scale radius). As already commented by PM07, the slightly worse behaviour obtained when the new force definition is used and $N_{ngbs} < 100$ can be ascribed to noise-induced gradients in the softening lengths altering the evaluation of the correction term; notwithstanding this, the ASEs for the ``adaptive+correction'' case are remarkably close to the minimum corresponding to the optimal choice of fixed softening throughout the full range of  $N_{ngbs}$. These results compare very well with those of PM07 (see their Fig.~$2$ and the second set of lines from the top).
\begin{figure}
\includegraphics[width=84mm]{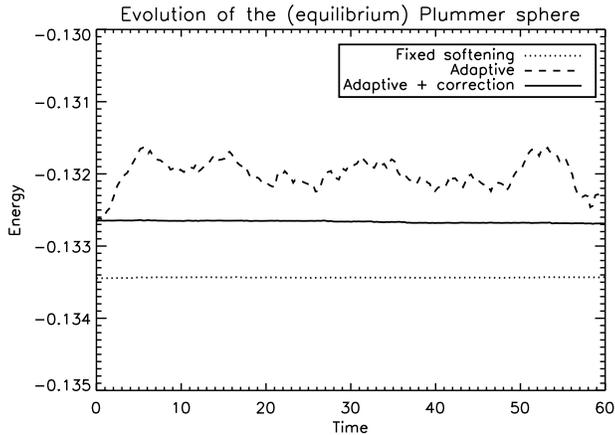}
 \caption{Behaviour of the total energy (kinetic plus potential) of the Plummer sphere as a function of time. The velocities were generated according to the equilibrium distribution function. For a description of the units, see the introduction to Sec.~\ref{sec:tests}. }
  \label{plummer_energy}
\end{figure}
\begin{figure}
\includegraphics[width=84mm]{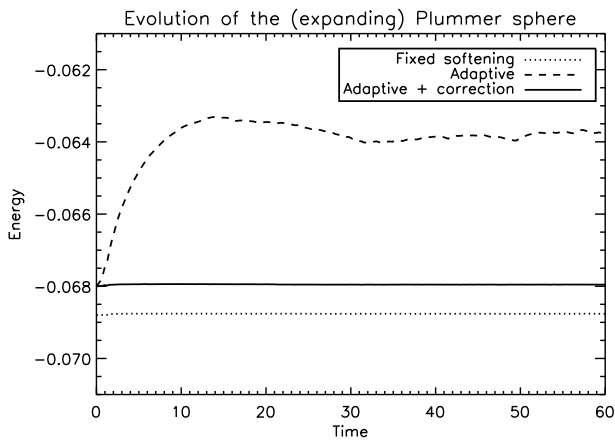}
 \caption{Behaviour of the total energy (kinetic plus potential) of the Plummer sphere as a function of time. The equilibrium velocity distribution has been multiplied by a factor $1.2$, leading to an initial global expansion of the system. For a description of the units, see the introduction to Sec.~\ref{sec:tests}.}
  \label{plummer_energy_pert}
\end{figure}
Choosing the optimal value for $\epsilon$ and $N_{ngbs}=60$ we have evolved the system in time and checked the conservation of the total (kinetic plus potential) energy. Fig. \ref{plummer_energy}  and \ref{plummer_energy_pert} show the results for two different initial velocity setups, corresponding to a dynamical equilibrium and a perturbed state respectively. In the first case we expect no major evolution of the system  (at least for several relaxation times) whereas in the second an initial overall expansion occurs before a state of equilibrium is reached. We refer again to PM07 and \citet{aarseth74} for details on the setup of the velocity profiles. The energy fluctuates or increases, reflecting the global changes in the softening lengths, when the adaptive formalism is used without changing the equation of motion accordingly; we register fluctuations of order per cent in the equilibrium set-up and an increase of $\approx 6\%$ in the perturbed set-up. Conservation is instead re-established down to time-stepping accuracy as soon as the correct equation of motion is used. Note that the initial total energies are different in the adaptive and fixed softening cases; this is due to the definition and evaluation of the potential energy being different in the two scenarios (see Sec.~\ref{sec:formalism}).\\
\subsection{Evolution of a Hernquist model}
\begin{figure}
\includegraphics[width=84mm]{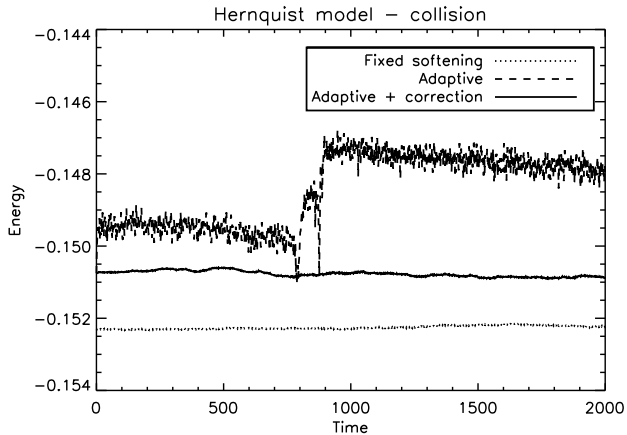}
 \caption{Behaviour of the total energy of two colliding Hernquist spheres as a function of time. The collision is head-on, with the two systems at an initial separation $ r = 150\;r_s$ and approaching each other at a speed $v = \sqrt{(2GM/r)}$. For a description of the units, see the introduction to Sec.~\ref{sec:tests}.}
  \label{hern_energy_merger}
\end{figure}

\begin{figure*}
\includegraphics{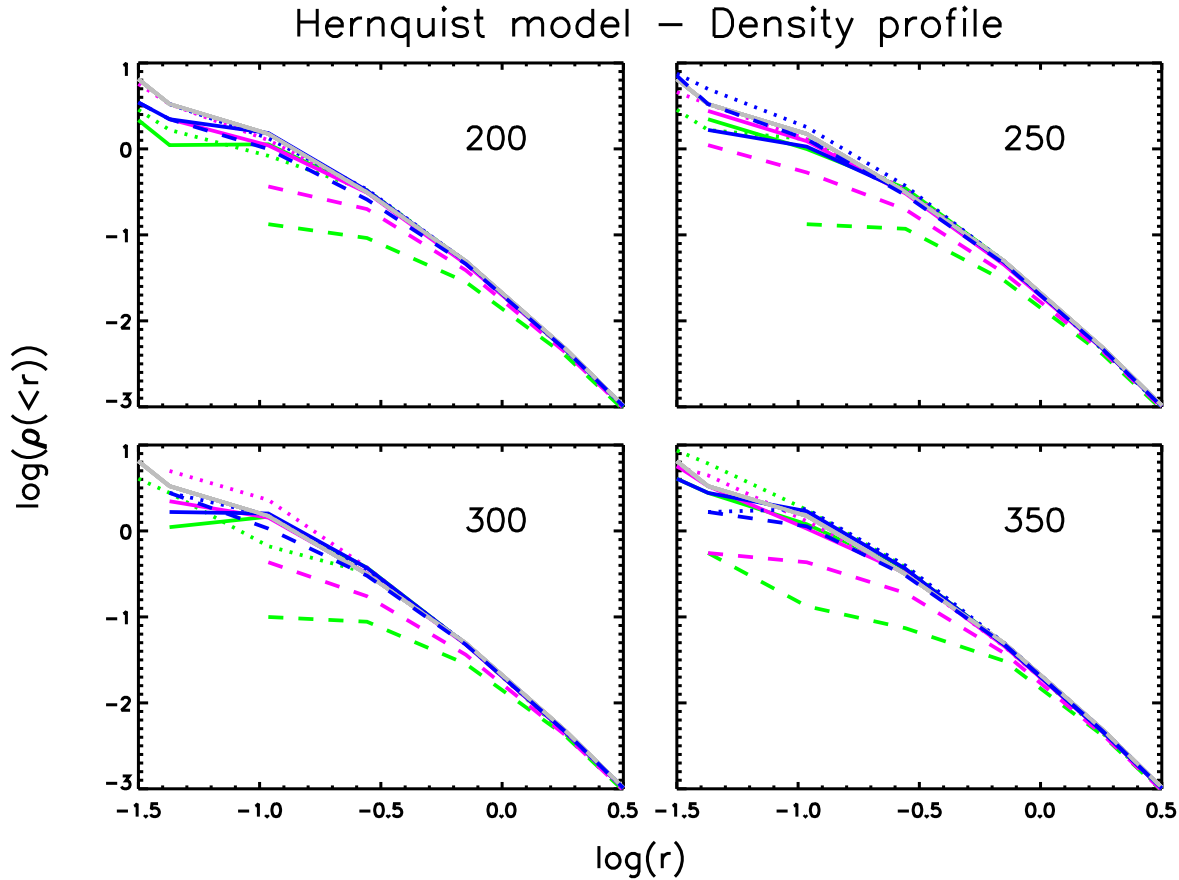}
 \caption{Cumulative density profile for the $1000$-particles Hernquist model at four different times ($t=200,250,300,350$). The gray lines represent the initial profile at $t=0$; solid lines correspond to the fiducial choice for $\epsilon (0.05)$ and $N_{ngbs} (60)$; dotted lines correspond to $\epsilon=0.001$ and $N_{ngbs}=30$, while dashed lines to $\epsilon=1$ and $N_{ngbs}=500$.}
  \label{hern_profile}
\end{figure*}

We now perform a similar analysis on a different density distribution; we generate a Monte Carlo realisation of a Hernquist model \citep{hernquist90}, whose density profile reads:
\begin{equation}
\rho(r) = \frac{GMr_s}{2\pi r(r_s + r)^{3}}.
\end{equation}
The model is cusped near the origin and provides a more realistic representation of the matter distribution in collapsed objects of cosmological interest, but the steep rise in the density profile makes the distribution more difficult to resolve than the Plummer case addressed in the previous section. We investigate the ASE and the energy evolution for a Hernquist model in equilibrium, sampled with $N=1000$ particles; as in the previous section, the total mass $M$ and the scale radius $r_s$ are set equal to unity.
\begin{figure}
\includegraphics[width=84mm]{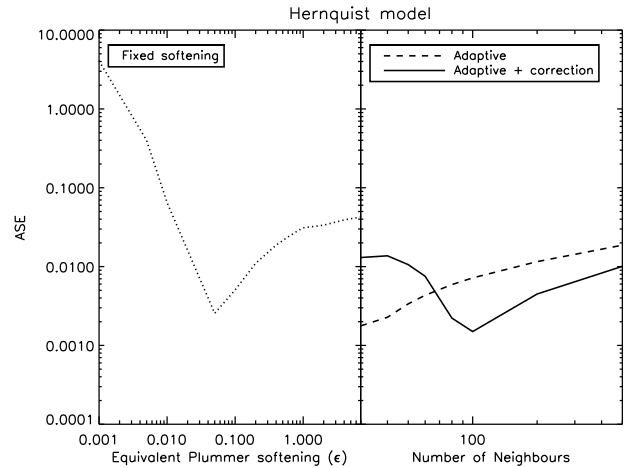}
 \caption{Average squared errors (ASE) of the force field generated by distributing $N=1000$ particles according to a Hernquist profile. The left panel shows the behaviour of the ASE as a function of the choice of (fixed) softening; the right panel shows the results of adaptive softening varying the number of neighbours.}
  \label{hern_ase}
\end{figure}
Fig.~\ref{hern_ase} shows the averaged square errors (ASE) of the simulated force field corresponding to different choices of both fixed and adaptive gravitational softening. Again we see how adapting softening provides relatively stable results when varying the number of neighbours from $30$ to $500$; conversely, one needs to choose the fixed softening accurately to avoid a strong misrepresentation of the force profile. We note that the optimal value for $\epsilon$ is smaller here than for the Plummer sphere ($\epsilon_{opt} \approx 0.05$ vs. $\epsilon_{opt} \approx 0.2$), a consequence of the different density distribution in the two models: the Hernquist being cusped and the Plummer being flat at the centre. A number of neighbours $N_{ngbs} \approx 60$ is enough to overcome the noise in the evaluation of the correction term and to provide better results than in the case where plain adaptive softening is used. For any number of neighbours, however, the force profile of the Hernquist model is reproduced better by adapting the gravitational softening. The results are in qualitative agreement with those of PM07 (see their Fig.~$3$ and the second set of lines from the top).
\begin{figure}
\includegraphics[width=84mm]{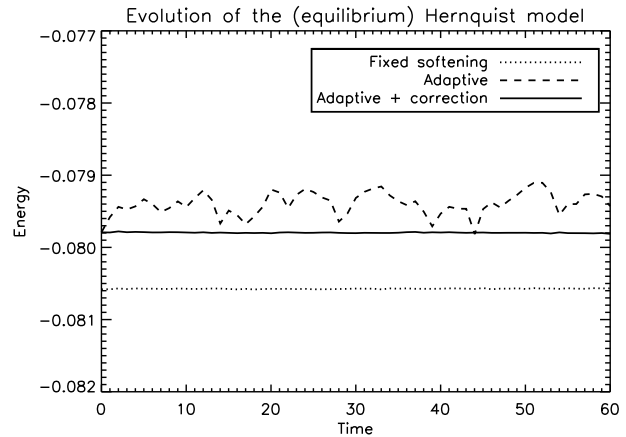}
 \caption{Behaviour of the total energy (kinetic plus potential) of the Hernquist model as a function of time. The velocities were generated according to the equilibrium distribution function. For a description of the units, see the introduction to Sec.~\ref{sec:tests}.}
  \label{hern_energy}
\end{figure}
As before, we now choose the optimal value for $\epsilon$ and $N_{ngbs}=60$ to evolve the system in time, checking the behaviour of the total energy. The results are shown in Fig.~\ref{hern_energy} for the equilibrium model; again, the energy fluctuates ($\approx 1\%)$ when the adaptive formalism is used without changing the equation of motion accordingly, whereas conservation is re-established down to time-stepping accuracy as soon as the correct equation is used.\\
A more demanding test for energy conservation is to follow the collision of two such particle distributions and check if the adaptive formalism maintains its conservative properties even in this more dynamically active scenario. To this purpose we have simulated the head-on collision of two different realisations of the Hernquist model, where the two systems were placed at a distance $r$ of $150$ scale radii and made approach each other at a speed $v = \sqrt{(2GM/r)}$. The evolution of the total energy is shown in Fig.~\ref{hern_energy_merger}; the collision between the two spheres at $t \simeq 900$ leaves a clear imprint in the total energy, when the softening is allowed to vary and the standard equation of motion is used, while as soon as the correction term is taken into account conservation is recovered.  \\
We now want to investigate whether or not the use of adaptive softening helps maintaining the original slope of the density profile in time. To this purpose we let the equilibrium system evolve for several dynamical times and compare the matter distribution among the different models. Fig.~\ref{hern_profile} shows the cumulative density profile at four different times, marked in the upper left corner. In all the panels the gray line represent the initial profile; the solid lines correspond to the optimal choice for fixed softening ($\epsilon_{opt} \approx 0.05$) and to $N_{ngbs} \approx 60$ for adaptive softening; the dotted lines correspond to $\epsilon=0.001$ and $N_{ngbs}=30$, whereas the dashed ones to  $\epsilon=1.0$ and $N_{ngbs}=500$. Again we see how sensitive the results are to the choice of $\epsilon$ and, correspondingly, how little they depend on $N_{ngbs}$, especially when the correct equation of motion is used. The density profile in the runs with adaptive softening is compatible with the original one at all times (and for all $N_{ngbs}$, when the correction term is added), with a tendency to outperform the results obtained when using fixed softening.
\subsection{Equilibrium structure of a polytrope}
Here we consider a system evolving under the action of both gravity and hydrodynamical forces. A homogeneous gas sphere of initial radius $r_0 = 1$ with equation of state $P=K\rho^{\gamma} \;,\gamma=5/3$ and initial null internal energy is released to the action of self-gravity and pressure force until hydrostatic equilibrium is reached. 
The process is speeded up by means of the standard SPH artificial viscosity together with a damping term in the force equation; their effect is to deliberately remove kinetic energy helping the system settling faster to equilibrium. An analytical solution for the internal structure of the system does not exist when $\gamma=5/3$; the exact radial density profile was then obtained by numerically integrating the corresponding Lane-Emden equation:
\begin{equation}\label{eq:lanemden}
\frac{\gamma K}{4\pi G(\gamma-1)}\frac{d^2}{dr^2}(r\rho^{\gamma-1}) + r\rho = 0.
\end{equation}
The reference for the setup and realisation of the test is Sec. 4.4 of PM07 and there we refer the reader for additional details on the generation of initial conditions and features of the run.\\
\begin{figure}
\includegraphics[width=84mm]{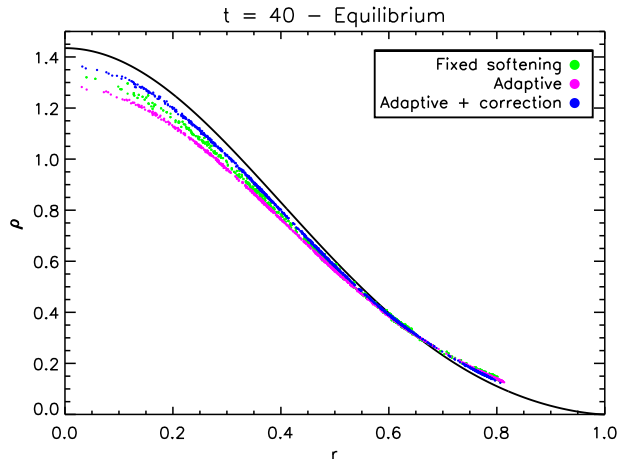}
 \caption{Radial density profile of a polytropic sphere in hydrostatic equilibrium. The solid black line represents the numerical solution of the Lane-Emden equation (Eq. \ref{eq:lanemden}); the points are the results obtained by evolving a $N=1445$-particles system with different choice of softenings.}
  \label{poly_prof}
\end{figure}
Fig.~\ref{poly_prof} shows the density profile of the simulated system at $t=40$, plotted against the equilibrium solution given by Eq.~\ref{eq:lanemden}. The units are as those introduced in the previous section; the time was chosen by following the oscillations in the density at $r=0.2$: when their amplitude reduced of a factor of $20$ with respect to the initial, we assumed equilibrium was reached.\\
Three cases are shown, namely fixed softening ($\epsilon \approx 1/40$ of the mean interparticle separation, as chosen by PM07) and adaptive softening ($N_{ngbs}=60$) with and without the use of the correction term. In all cases we have used $60$ neighbours for the SPH computations. As found by PM07, the use of adaptive softening provides a better representation of the density profile, especially the inner regions; the higher central densities can be related to a smaller gravitational softening than in the fixed case (at least when the correction term is employed), while the slightly better agreement at large radii comes from much larger softenings reducing the noise of the particle distribution.\\
\begin{figure}
\includegraphics[width=84mm]{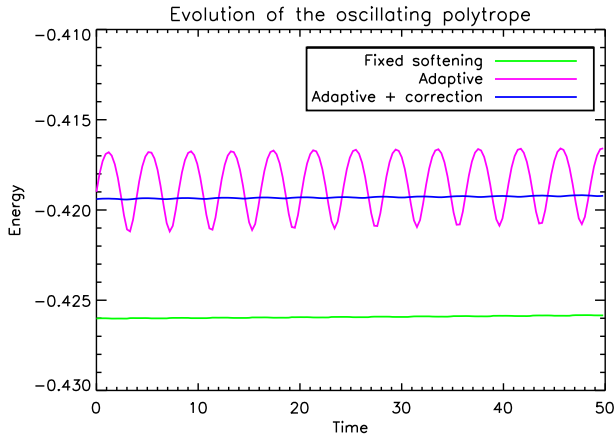}
 \caption{Behaviour of the total energy (kinetic, potential and internal) of the polytropic sphere as a function of time. The system has been perturbed from the equilibrium state at $t=40$ by the induction of radial oscillations. For a description of the units, see the introduction to Sec.~\ref{sec:tests}. }
  \label{poly_ene}
\end{figure}
The equilibrium solution just found has then been perturbed by inducing radial oscillations on the sphere. The system was let evolve under gravity and pressure force only, the behaviour of its total energy being now of interest. The results are displayed in Fig. \ref{poly_ene}. The system oscillates with a period $\tau \approx 4$, close to the expected $3.82$ for the fundamental mode of oscillation of such a polytrope. This is seen clearly in the behaviour of the energy when adaptive softening is used and the correction to the equation of motion is neglected; the evaluated potential becomes cyclically deeper and shallower following, respectively, the overall decrease and increase of the softening lengths. Energy conservation is re-established at the level of the runs with fixed softening once the system is evolved according to the appropriate equations.

\section[]{Performance in a cosmological environment}
\label{sec:cosmo}
The problem of the formation and evolution of the large-scale structure of the Universe has no analytical solution and this complicates the analysis when different simulations of the same system are compared; to assess the results and determine which technique is dealing best with the problem is not obvious anymore. Increasing the resolution of a cosmological simulation generally goes along with extending the representation of the initial perturbation field to smaller scales, resulting in new fluctuations entering the initial conditions and therefore, strictly speaking, in a new system. In order to address the problem of structure formation and of how this is dealt with by the different softening formulations we have decided to perform a series of simulations sharing the same initial power spectrum of fluctuations, but differing in the total number of particles (\citealt*{binney2004}): the same structures form at all the resolution levels, but are more accurately described as more and more particles populate them. If the use of adaptive softening allowed to anticipate the behaviour shown by the standard runs at higher resolution, this would assess the superiority of the method over the usual choice of fixed softening. \\ 
The simulated system here consists of a $100\; h^{-1}\mathrm{Mpc}$ side periodic box in a $\Lambda$CDM cosmology defined by the following choice of parameters:
\begin{eqnarray}
 & & \Omega_{\rm tot} = \! 1, \; \Omega_m = \!0.3, \; \Omega_b=0.04, \; \Omega_{\Lambda}=0.7, \nonumber \\
 & & h = 0.7, \;\sigma_8=0.9, \; n_s=1\,,
\end{eqnarray}
where $h$ and $\sigma_8$ are the values, at redshift zero, of the dimensionless Hubble parameter and rms of the mass fluctuations smoothed on a scale of $8\; h^{-1}\mathrm{Mpc}$, whereas $n_s$ is the index of the primordial spectrum of fluctuations. 
Two sets of simulations have been performed: three runs with $64^3$, $128^3$ and $256^3$ particles using fixed softening and two runs with  $64^3$ and $128^3$ particles using adaptive softening (with and without the correction term). The initial conditions were generated at $z=60$ and differ among them only in the number of particles resolving the same initial fluctuation field. The pre-initial uniform distribution, represented by an equally-spaced grid of particles, is applied a displacement field generated on $64^3$, $128^3$ and $256^3$ grids, depending on the resolution level. The power spectra are anyway identical in all the three cases and they are truncated at the smallest frequency resolved on a $64^3$ grid. The simulations follow only the evolution of the dark matter component of the density field and the resulting mass associated to a single particle varies from $\approx 32$ to $4$ and $0.5 \cdot 10^{10}\;h^{-1}\mathrm{M}_{\odot}$ depending on the resolution level of the run. In all the cases we have used the TreePM algorithm varying the grid from $64^3$ to $128^3$ and $256^3$, according to the resolution level. The features of the simulations are summarised in Table~\ref{table}.\\
\begin{table*}
 \centering
 \begin{minipage}{140mm}
  \caption{Basic features of the simulations presented in Sec.~\ref{sec:cosmo}.}
  \label{table}
\begin{tabular}{l*{2}{c}r*{2}{c}}
\hline
Name                    & $\epsilon$    & $N_{ngbs}$     & $m_p$                & $k_{min}$           & $k_{max}$        \\
                        & $[h^{-1}kpc]$ &               & $[h^{-1}M_{\odot}]$  & $[h\;kpc^{-1}]$      & $[h\;kpc^{-1}]$   \\
\hline
Fixed - $64^3$     	& $39$  	& -             & $31.76\cdot10^{10}$  & $6.28\cdot10^{-5}$  &$2\cdot10^{-3}$  \\
Fixed - $128^3$    	& $19.5$        & -             & $3.97\cdot10^{10}$   & $6.28\cdot10^{-5}$  &$2\cdot10^{-3}$  \\              
Fixed - $256^3$    	& $10$          & -             & $0.5\cdot10^{10}$    & $6.28\cdot10^{-5}$  &$2\cdot10^{-3}$  \\             
Adapt - $64^3$     	& -             & $60 \pm 0.1$  & $31.76\cdot10^{10}$  & $6.28\cdot10^{-5}$  &$2\cdot10^{-3}$  \\              
Adapt+corr - $64^3$	& -             & $60 \pm 0.1$  & $31.76\cdot10^{10}$  & $6.28\cdot10^{-5}$  &$2\cdot10^{-3}$  \\             
Adapt - $128^3$    	& -             & $60 \pm 0.1$  & $3.97\cdot10^{10}$   & $6.28\cdot10^{-5}$  &$2\cdot10^{-3}$  \\            
Adapt+corr - $128^3$ 	& -             & $60 \pm 0.1$  & $3.97\cdot10^{10}$   & $6.28\cdot10^{-5}$  &$2\cdot10^{-3}$  \\              
\hline
\end{tabular}
\end{minipage}
\end{table*}
The value of fixed softening was chosen to be $1/40$ of the mean interparticle separation in the system, i.e. the smallest advisable value according to the generally accepted criterion that limits this scale to avoid too negative binding energies in close pairs. The choice of the number of neighbours for the adaptive runs corresponds to the minimum number ensuring a robust evaluation of the correction term. To this purpose we have monitored the changes in the two crucial quantities entering the evaluation of the correction term, namely $\zeta$ and $\Omega$ (Eq. \ref{eq:zeta} and \ref{eq:omega}, respectively), when varying $N_{ngbs}$. The code was run on the $z=0$ snapshot of the $128^3$ simulation employing fixed softening with  $N_{ngbs}$ varying from $16$ to $80$ with a step $\Delta N_{ngbs}=8$. 
\begin{figure}
\includegraphics[width=84mm]{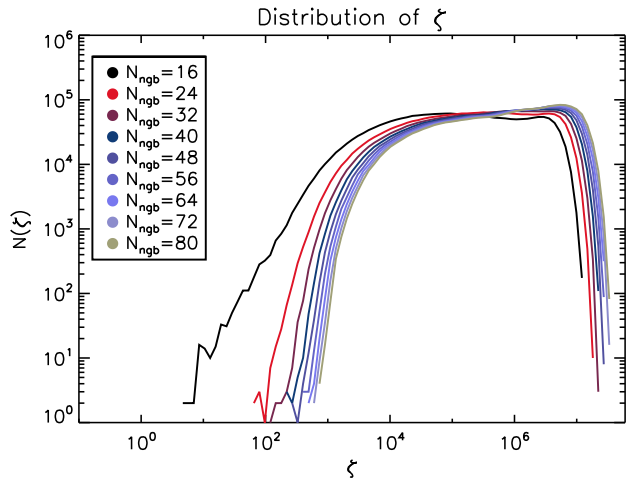}
 \caption{Number of particle with  $\zeta$ (Eq.~\ref{eq:zeta}) in a certain range as a function of $\zeta$. The code was run on the same clustered distribution of particles using different values for the number of neighbours.}
  \label{zeta_distr}
\end{figure}
\begin{figure}
\includegraphics[width=84mm]{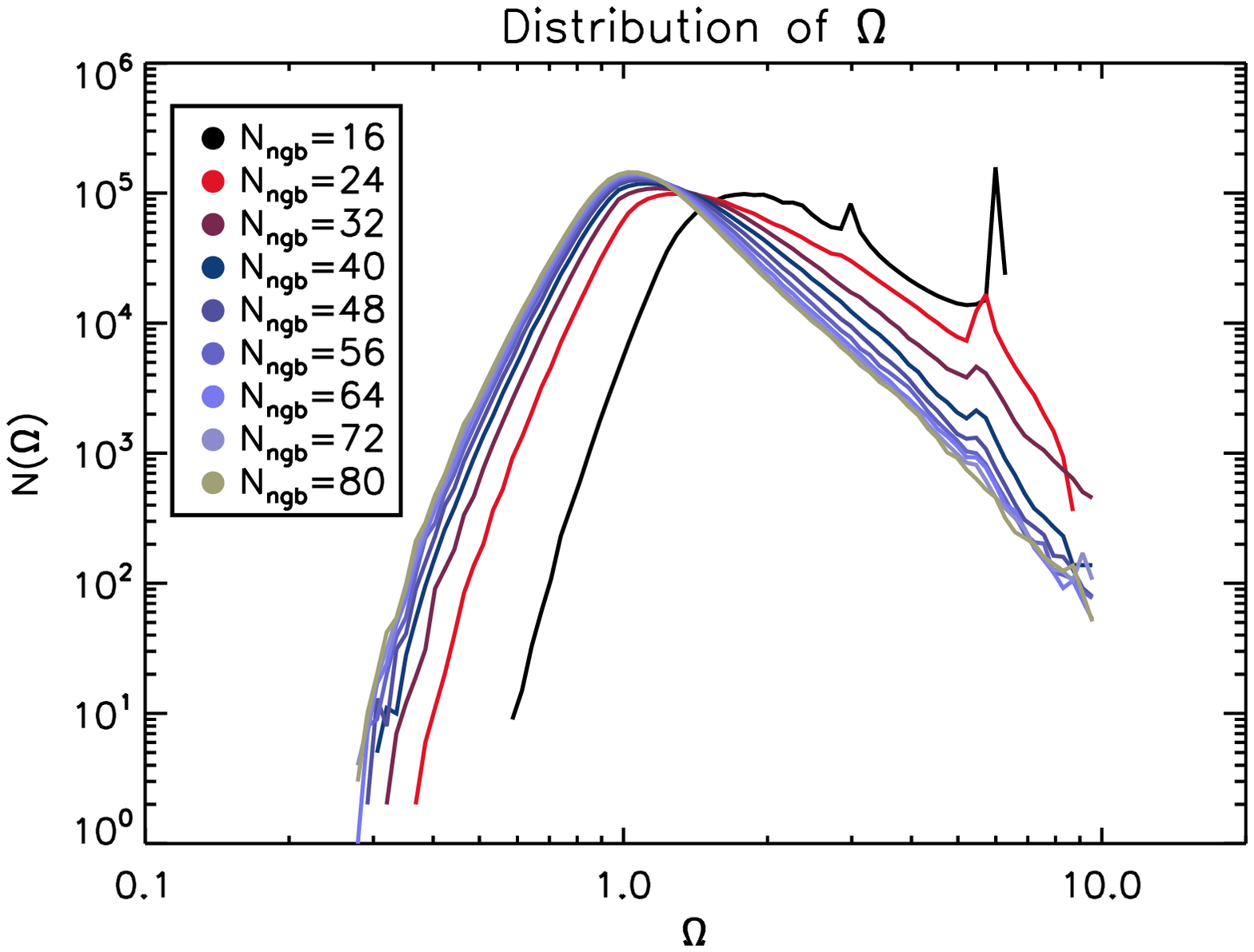}
 \caption{Number of particle with  $\Omega$ (Eq.~\ref{eq:omega}) in a certain range as a function of $\Omega$. The code was run on the same clustered distribution of particles using different values for the number of neighbours.}
  \label{omega_distr}
\end{figure}
Fig. \ref{zeta_distr} and \ref{omega_distr} show the distribution of the two quantities when varying $N_{ngbs}$; we can conclude that a number $N_{ngbs}= 60$ is enough to ensure a converged evaluation of $\zeta$ and $\Omega$ and that the resulting correction term would be robustly estimated. We have also run the simulations using adaptive softening and no correction to the equations of motion, using the same number of neighbours as the fully adaptive ones; although the results of the previous section suggest that the introduction of the correction term is crucial to the proper functioning of the adaptive formalism, we will anyway show the behaviour of these simulations for completeness. \\
In the following subsections we summarise the main results.

\subsection{Global behaviour I - Clustering}
\begin{figure*}
\includegraphics[]{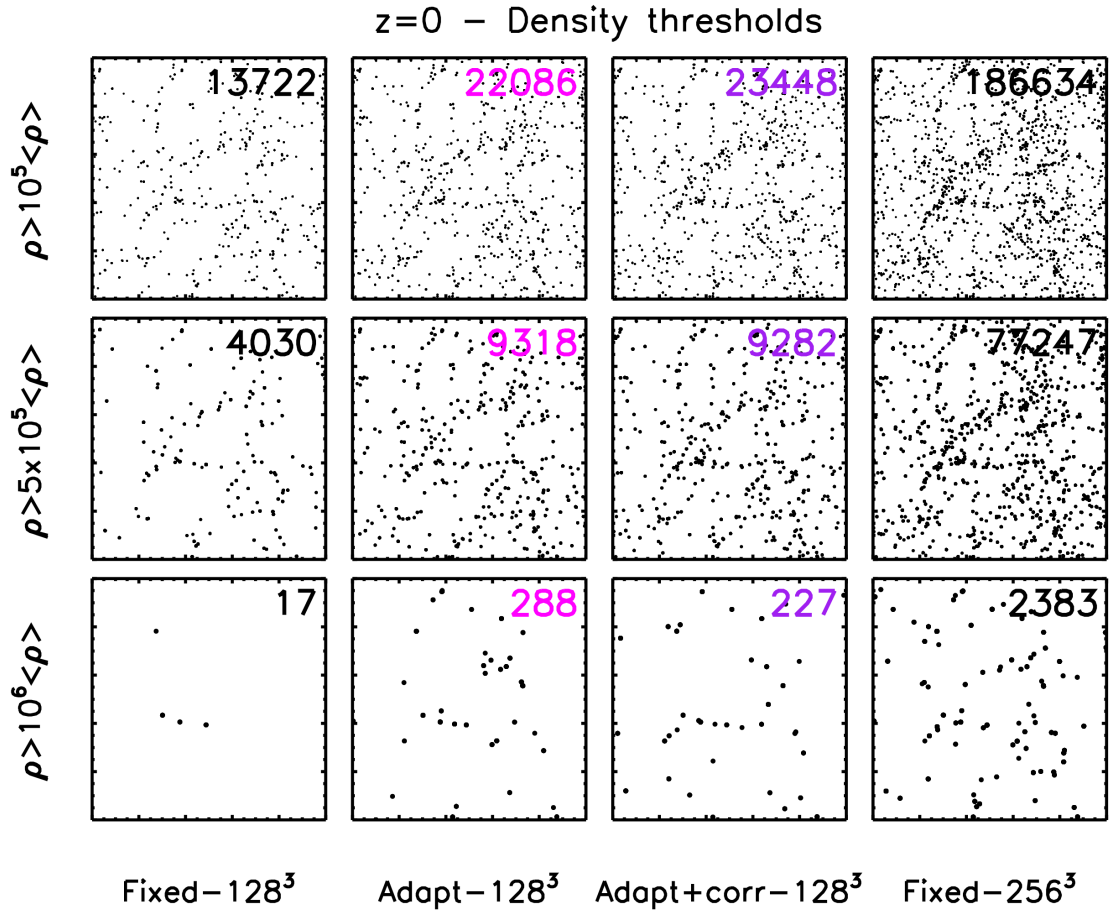}
 \caption{Particles with an associated density greater than a threshold. Their total number is shown in the upper-right corner. The full $100\; h^{-1}\mathrm{Mpc}$ side box is shown and all the particles are plotted; some are not distinguishable from their neighbours due to their proximity, which in some cases can reach $0.5 \; h^{-1}\mathrm{kpc}$. The densities were computed as the position-weighted sum of the nearest neighbours, in the usual SPH fashion; the same number of neighbours ($60$) has been used for all the simulations. The results are shown at $z=0$.} 
  \label{dens_thres}
\end{figure*}

The level of clustering in the simulations has been first assessed qualitatively by investigating the densities achieved throughout the box. Fig.~\ref{dens_thres} shows the particles with an associated density greater than $10^5$, $5 \cdot 10^5$ and $10^6$ times the average density; the results are from the redshift zero snapshots of the $128^3$ and $256^3$ simulations. The densities have been computed in the SPH fashion, using $60$ neighbours for all the runs.
As one can see, the runs with adaptive softening reach higher particles densities. Noticeably, the regions where this enhancement of clustering is observed correspond to the high density regions of the $256^3$ simulation; indeed, moving from left to right (i.e. from the fixed softening case to the fully adaptive one) the same areas can be seen to be more and more populated. The numbers in the upper-right corner corresponds to the total number of particles surviving the density threshold; interestingly, the number of these high-density particles in the $256^3$ simulation correspond to a mass of $\approx 23000,\; 10000,\; 300$ particles at the $128^3$ resolution level.\\
On a more quantitative level, we have also analysed the clustering by means of the two-point correlation function $\xi(r)$. This statistics represents the excess probability, compared to a uniform random distribution, of finding pairs of particles at a given spatial separation. 
\begin{figure}
\includegraphics[width=84mm]{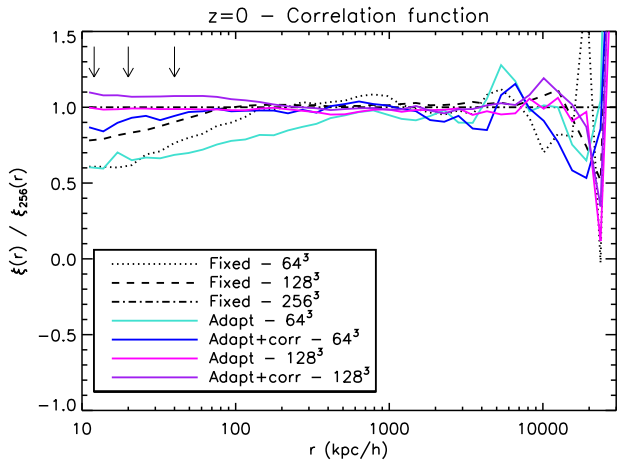}
 \caption{Two-point correlation function at $z=0$ for a series of simulations of a $100\; h^{-1}\mathrm{Mpc}$ side box in a $\Lambda$CDM universe; shown is the ratio to the results of the $256^3$ simulation. The runs differ only in the choice of softening and in the total number of particles used to sample the initial perturbation field. The arrows indicate the equivalent-Plummer values of the gravitational softening in the standard runs.}
  \label{conv_corr_fct}
\end{figure}
Fig.~\ref{conv_corr_fct} shows the ratio of the correlation functions obtained in the different runs to the result of the  $256^3$ simulation; the curves are shown at $z=0$ and starting from separations of $10 \; h^{-1}\mathrm{kpc}$, which correspond roughly to the resolution scale at the $256^3$ level. Other than for some discrepancies at separation of several megaparsecs\footnote{These discrepancies have no physical meaning and are due to a somewhat less accurate determination of the correlation function at large separations. The method we use is Monte-Carlo based and tuned to provide very solid results at small separations, at the expense of accuracy on large scales. We checked that the differences we register are within the typical uncertainties due to the method when the correlation function is computed at these separations.}, the correlation functions of the three runs with fixed softening overlap almost perfectly down to $100\; h^{-1}\mathrm{kpc}$. Below that scale the differences in resolution result in a progressively larger amplitude. Since the behaviour of $\xi(r)$ at the small-$r$ end is mainly determined by the distribution of particles within individual halos, we could conclude that the internal distribution of particles in a structure becomes denser as the resolution of the simulation is increased. The runs with adaptive softening produce a correlation function which is in full agreement with that of the ``fixed'' run at the same resolution down to roughly $100\; h^{-1}\mathrm{kpc}$; at smaller separation the amplitude grows instead larger, approaching the results obtained at higher resolution when using fixed softening. This is particularly evident in the $64^3$-run using adaptive softening and the correction of the equation of motion (``Adapt+corr - $64^3$'', blue curve): not only is the amplitude at small separation ($10$ - $50\; h^{-1}\mathrm{kpc}$) higher than the other two $64^3$ runs, but the behaviour at scales  between $50$ and $200\; h^{-1}\mathrm{kpc}$ is indistinguishable from the reference $128^3$-run (black, dashed curve). Both the ``adaptive'' runs at the $128^3$ resolution level behave well in reproducing the correlation function of the $256^3$ simulation, the one with correction (purple curve) even slightly exceeding the results for the $256^3$ case on scales below $100\; h^{-1}\mathrm{kpc}$.

\subsection{Global behaviour II - Mass function}
The search for structures in the simulations has been carried out with the algorithms \fof  (\citealt{fof}) and \subf (\citealt{springel01a}). Structures are first identified as collections of $N > N_{min}$ particles separated by mutual distances smaller than some fraction $b$ of the mean interparticle separation. These so-called  ``FOF halos'' are later examined by \subf for the identification of self-bound substructures and the removal of spurious background particles. In the simulations presented here the halos have been searched using a linking length $b=0.16$ and a minimum threshold of $32$ particles. The internal structure of these candidate objects has then been probed in order to identify local, gravitationally-bound overdensities; those containing at least $20$ particles were addressed to as ``subhalos'' leaving the others as part of the smooth halo component. Finally, particles not gravitationally bound to any substructure of the parent FOF halo were dismissed. Note that \subf was modified in the unbinding part so that the evaluation of the gravitational potential takes into account the individual softening length of the particles.
\begin{figure}
\includegraphics[width=84mm]{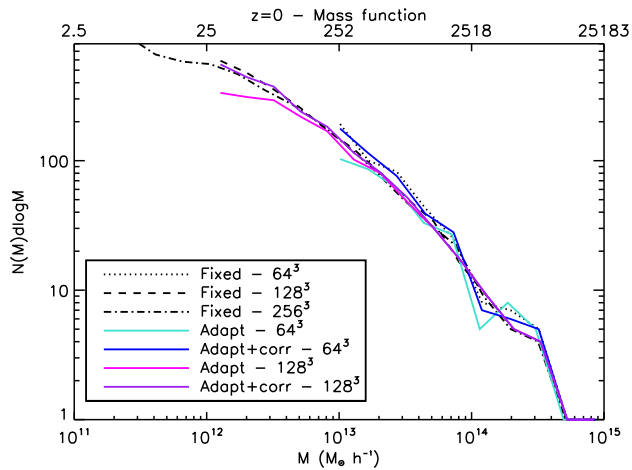}
 \caption{FOF mass function at $z=0$ for a series of simulations of a $100\; h^{-1}\mathrm{Mpc}$ side box in a $\Lambda$CDM universe. The upper x-axis displays the number of particles in the $128^3$ run corresponding to the masses in the lower axis. The mass functions are truncated at a mass corresponding to $32$ particle masses in every resolution level (i.e. at $\approx 10^{13},\; 1.3\cdot 10^{12},\; 1.7\cdot 10^{11} h^{-1}\mathrm{M}_{\odot}$ for the $64^3$,$128^3$ and $256^3$ run, respectively). The simulations differ only in the choice of softening and in the total number of particles used to sample the initial perturbation field. }
  \label{conv_mfct}
\end{figure}
Fig.~\ref{conv_mfct} shows the mass function of the \fof halos at $z=0$; the number of objects per logarithmic mass interval is plotted for the different runs down to the mass limit corresponding to $32$ particles. When the correct equation of motion is used, the agreement between the simulations with adaptive and fixed softening at the same resolution level is striking throughout the full mass range; when the correction term is instead ignored, the mass functions drop visibly at low masses, the number of objects containing a number of particles less than of order $N_{ngbs}$ being significantly underestimated. Overall, the runs with adaptive softening and correction underestimate the total number of FOF halos by $\simeq 3\%$, as opposed to a $\simeq 25\%$ for those without the correction term; if we considered only halos containing more than $100$ particles, these percentages would fall to $\simeq 1\%$ and $\simeq 14\%$, respectively. Similar results are obtained when considering the ``spherical overdensity'' masses $M_{\Delta}$\footnote{These are defined as the masses of the spherical regions centred on the potential minimum of the smooth halo and corresponding to an overdensity of $\Delta$, typically $\approx 200$, with respect to either the critical density $\rho_{crit}$ or to the mean background density $\rho_{m} = \Omega_m\rho_{crit}$. Hereafter, when referring to either the virial radius or the virial mass of an object, we assume the overdensity to be defined with respect to the mean background density.}; this hints to the fact that the global shape of the structures in the simulations is not changed significantly by the adoption of the adaptive softening formalism. \\
The upturn in the curve for the $256^3$ simulation at the low-mass end resembles the effect of spurious halos seen in warm dark matter simulations (see, e.g. \cite{wang07} and their Fig.~$9$); these objects form from the artificial fragmentation of filaments and have masses much smaller than the free-streaming mass of the model being adopted, their origin being entirely due to the specifics of the pre-initial conditions. Having our simulations a truncated power-spectrum, we might in principle be hitting the same problem and have a contribution to our mass functions coming from such spurious structures. This could indeed be the case for the $256^3$ simulation at the very low-mass end, a regime we are not interested in for our comparison. As for the $128^3$ simulations, the problem would affect objects with masses corresponding to a handful of particles that do not survive the $32$-particles limit and therefore do not enter the evaluation of the mass functions. We also specify that this problem does not invalidate the results we displayed in the previous section in terms of correlation functions; the number of particles in the alleged spurious structures in the $256^3$ simulation is insufficient to affect the evaluation of the correlation function at the scales we are interested.

\subsection{Internal halo properties} 
\begin{figure}
\includegraphics[width=84mm]{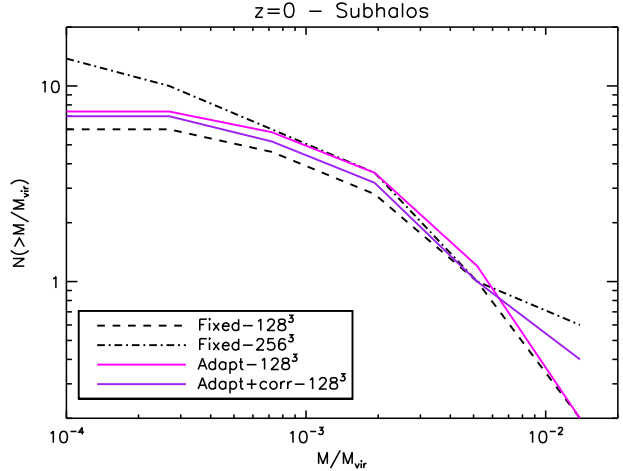}
 \caption{Average subhalo mass function at $z=0$ for the five most massive halos in the simulations. The subhalo masses have been normalised to the virial mass of the host and vary from $8\cdot 10^{11}$ to  $4\cdot 10^{13} h^{-1}\mathrm{M}_{\odot}$, corresponding to $\approx 20$ and $1000$ particles, respectively.}
  \label{shalos}
\end{figure}

The results in terms of correlation function and mass function suggest
that the main differences between the runs are likely to manifest in
the internal structures of the collapsed objects. We will investigate
here what effects the different definitions of softening have in the
representation of the most massive halo formed in the
simulations. 
\begin{figure}
\includegraphics[width=84mm]{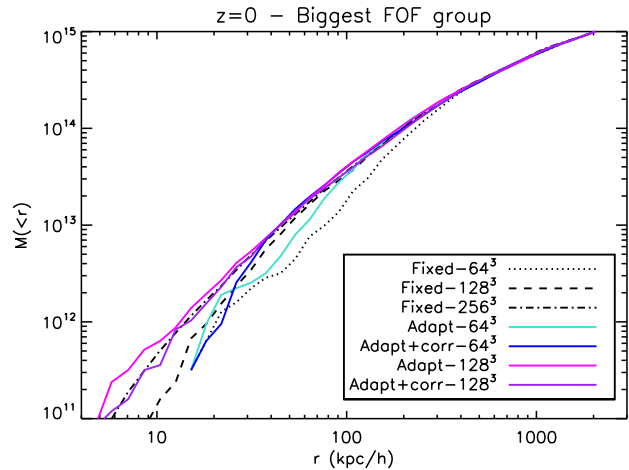}
 \caption{Cumulative mass distribution for the most massive FOF group at $z=0$.}
  \label{conv_prof}
\end{figure}
Fig.~\ref{conv_prof} shows the cumulative mass distribution in the halo out to the virial radius. The values differ by fractions of per cent in the various runs and assess around $2.4\; h^{-1}\mathrm{Mpc}$. Again it is evident how the use of adaptive softening enhances the clustering of particles anticipating the behaviour of higher resolution simulations. The results of the two runs with adaptive softening and correction (``Adapt+corr - $64^3$'', blue curve; ``Adapt+corr - $128^3$'', purple curve) are particularly worth of notice: the first reproduces the higher-resolution results down to $20\; h^{-1}\mathrm{kpc}$, whereas the second is perfectly compatible with the highest resolution run down to less than $5\; h^{-1}\mathrm{kpc}$. Another way to interpret this result is in terms of mean inner density as a function of the enclosed number of particles. As noted by \citet{moore1998} and \citet{power2003}, obtaining robust results in regions closer to the centre, where the density is higher, demands increasingly large particle numbers. As in Fig. $14$ of \citet{power2003} we have compared the mean inner density measured at different fractions of the virial radius as a function of the enclosed particle number for all the runs. The results are displayed in Fig.~\ref{power14}.
\begin{figure}
\includegraphics[width=84mm]{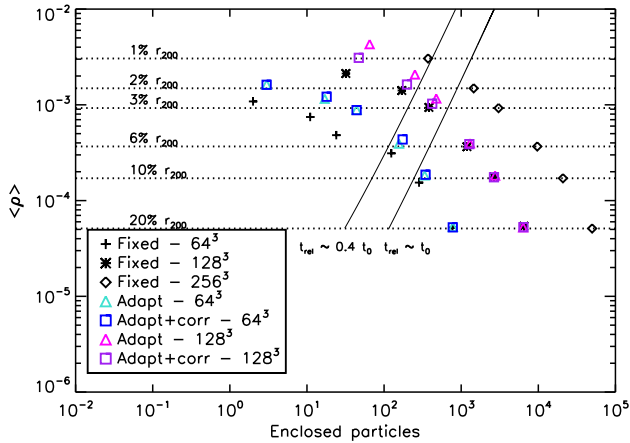}
 \caption{Mean inner density as a function of the enclosed number of particle at different fractions of the virial radius. The halo under consideration is the same as in Fig. \ref{conv_prof}. The solid lines separate the regions, at their right, where the average collisional relaxation time ($t_{rel}$) exceeds some fraction of the age of Universe ($t_0$).}
  \label{power14}
\end{figure}
The $64^3$ simulation using fixed softening provides converged results at most down to $\simeq 6\%$ of the virial radius. At smaller radii the results start to diverge from those obtained at higher resolution and cannot be trusted anymore. The two runs with adaptive softening behave instead very well down to $3\%$ of the virial radius. Similarly, the $128^3$ simulation is reliable down to $\simeq 2\%$ of the virial radius when using fixed softening and down to $1\%$ of the virial radius when adaptive softening is employed. \citet{power2003} relate the converged side of the enclosed-$\rho$ vs. enclosed-N plot to the regions where the average collisional relaxation time $t_{rel}$ exceeds some fraction of the age of Universe $t_0$ (between $0.6$ and $1$). In our case, depending on whether we consider the results at $6\%$ of the virial radius converged or not, we could extend the regions down to the radii where $t_{rel} \approx 0.4 t_0$. In any case, it is not clear whether the $256^3$ run can be considered converged according to this criterion. It is anyway worth of notice that the $128^3$ adaptive run with correction reproduces perfectly the enclosed density of the $256^3$ at  $1\%$ of the virial radius.\\
Similar results hold when investigating the properties of substructures. Fig.~\ref{shalos} shows the number of subhalos with mass greater than a certain value; the curves represent the average over the five biggest halos in the simulations and the masses have all been normalised to the virial mass of the host. Due to the poverty of substructures we are not showing the results for the lowest resolution simulations and we are in general limited to qualitative considerations. As can immediately be seen though, the runs with adaptive softening lie closer to the higher resolution simulation than does the run with fixed softening. 

\subsection{Comments}
\begin{figure}
\includegraphics[width=84mm]{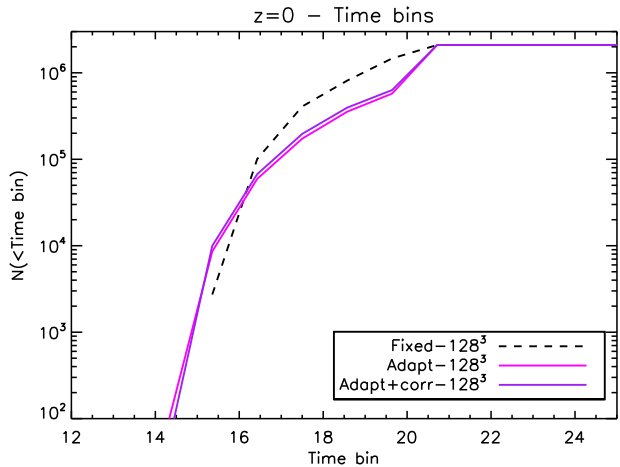}
 \caption{Cumulative distribution of particles in time bins at $z=0$. The simulations were run using $29$ time bins. Smaller numbers corresponds to finer intervals. }
  \label{timebin}
\end{figure}

\begin{figure}
\includegraphics[width=84mm]{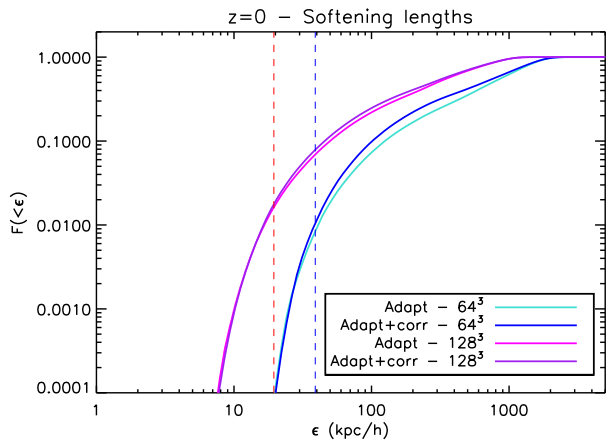}
 \caption{Cumulative distribution of softenings at $z=0$. The softenings are expressed in terms of equivalent-Plummer values ($\epsilon$). The red (blue) dashed curve represents the value of the gravitational softening in the $128^3$ ($64^3$) ``fixed'' simulations.}
  \label{soft}
\end{figure}

All the results shown so far hold at redshift zero. As shown in Fig.~\ref{soft}, even at this redshift only $\approx 1 \%$ of the particles have softening smaller than $1/40$ of the mean interparticle separation and at higher redshifts this behaviour becomes even more extreme. This has the effect of reducing the amplitude of the correlation function more rapidly than in the standard runs when going back in time. A more quantitative representation of this effect will be given in the next section (Fig.~\ref{mmII_corr}). The mass functions are not substantially affected though, at least not for objects more massive than the $32$-particles limit (Fig.~\ref{mmII_mfct}).\\
As the particle timesteps depends on the value of the gravitational
softening (see Eq.~\ref{eq:timecriterion}), it is natural to expect
that they would now span a wider range of values. Fig.~\ref{timebin}
shows that this is the case; plotted are the distributions of
particles in time bins\footnote{In {\sc GADGET} the simulated timespan
  is mapped onto the integer interval $[0, 2^{N_{timebins}}]$. This
  interval is split recursively into timebins, the smallest of which
  has length $2$. Each time, the code computes individual timesteps
  for the particles and distributes them in the correspondent bin; the
  smallest populated bin sets the next timestep for the simulation.}
at redshift zero for the three $128^3$ simulations: the ``adaptive''
runs tend to have particles in lower bins than in the ``fixed''
simulation and, at the same time, more particles in the highest
bin. This is a consequence of what was just mentioned, namely that only a very small fraction of the particles obtains an adaptive softening smaller than the fiducial choice for the standard simulations at the same resolution level. As finer time bins get populated, the overall number of timesteps increases when adaptive softening is used (along with the number of operations performed within them); at the same time, since the highest bins become more crowded, the average number of active particles \-- i.e. those that are advanced in a timestep \-- decreases instead, almost compensating for this overhead. 
Increasing $N_{ngbs}$ has the effect that one intuitively imagines, namely that of increasing the softening associated to the particles; the differences again stand out more evident when looking at the correlation functions, whose amplitude at small separations reduces, whereas the mass functions are confirmed to be considerably less sensitive to variations in $N_{ngbs}$.\\
All the simulations discussed in this section the gravitational softening was prevented to fall below $0.1\%$ of the TreePM splitting scale $r_s$ (corresponding to $558\;pc$ for the $64^3$ case and to $279\;pc$ for the $128^3$ one; in both cases, this translates to $\approx 1.5\%$ of the softening scale in the ``fixed'' runs); this implies that the maximum density achievable in the simulation corresponds to roughly $10^9 \bar{\rho}$. As pointed out in Sec.~\ref{sec:gadget}, the presence of a lower bound is not crucial and it is introduced mainly to prevent few particles in over-dense regions from slowing down the simulation. No upper limit to the gravitational softening was set in these simulations; we do not find substantial differences between the runs with and without such a constraint, especially if the correct equation of motion is used. We have also checked both sets against Tree-only runs and found perfectly compatible results.\\
Although we both observe an increased level of clustering in massive objects, our global results in terms of
particle correlation function and halo mass function somewhat differ from those of BK09; when comparing the results from the adaptive runs to the standard case with fixed softening, BK09 register a deficit in the number of small-mass objects and a slight lowering in the amplitude of the two-point correlation function at small separation, whereas we notice no change in the mass function and instead an overall enhanced level of particle clustering.
The simulated background cosmology differ, but this should have no effect in the analysis we are interested here, which is focused on relative differences between the runs more than on absolute results of cosmological interest. The larger linking length we have employed is not at the origin of the discrepancy between the FOF mass functions, nor we think is the different evaluation of the correlation function responsible for the antithetical outcomes.\\
The results differ also on the more technical timing level; in our simulations we do not register substantial saving of computing time when adaptive softening is used, but we do notice increasingly better performances when the number of neighbours is increased. BK09 halve the wallclock time of the reference run when using adaptive softening with $32$ neighbours and leave it unchanged when using $48$. If we set an upper limit to the softening of order the splitting scale $r_s$ (as in BK09), we progressively cut the timing when going from $32$ up to $60$ neighbours and leave it essentially unaltered by increasing the number even further; the minimum wallclock time is reached for $N_{ngbs} \simeq 60$ and it equals that of the simulation with fixed softening. If no upper limit is set for the softening, the wallclock times tend to increase; a minimum is reached for a number of neighbours $\simeq 40$ and it is around $\sim 30\%$ higher than in the standard run.\\
We ascribe the responsibility for the different behaviour of the implementation in the various aspects to the respective mother codes. 

\section{Simulating a mini version of the Millennium-II}
\label{sec:mmII}
\begin{figure}
\includegraphics[width=84mm]{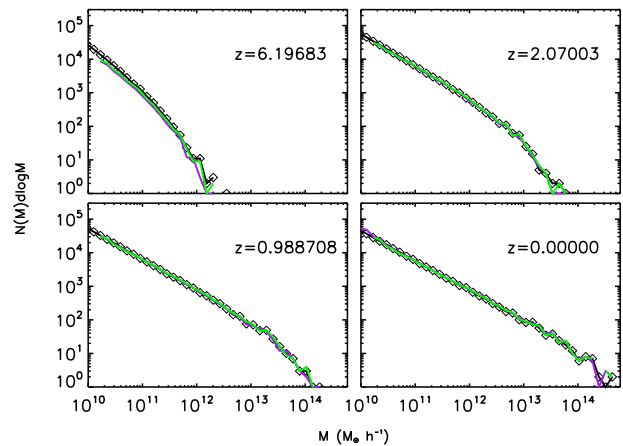}
 \caption{FOF mass functions from the mII (black curve, diamonds), mmII (green curve) and ``adaptive''-mmII (purple curve) simulations at different redshifts. See the text for a description of the three cases.}
  \label{mmII_mfct}
\end{figure}

As a more demanding application, we have investigated the effect of adaptive softening on a cosmological simulation whose small-scale clustering is known in detail by an extremely high resolution run. We have used the modified \gadgt~on the initial conditions of the ``mini''-Millennium-II simulation (hereafter mmII), a low-resolution version of the better known Millennium-II (hereafter mII, \citealt{mII}). The mmII follows the evolution of $432^3$ particles (as opposed to $2160^3$ particles for the mII simulation) within a box of side $100\; h^{-1}\mathrm{Mpc}$. The underlying cosmology is a $\Lambda$CDM with parameters
\begin{eqnarray}
 & & \Omega_{\rm tot} = \! 1, \; \Omega_m = \!0.25, \; \Omega_b=0.045, \; \Omega_{\Lambda}=0.75, \nonumber \\
 & & h = 0.73, \;\sigma_8=0.9, \; n_s=1\,.
\end{eqnarray}
This results in the particles having masses $m=8.61 \cdot 10^{10}\;h^{-1}\mathrm{M}_{\odot}$ , a factor $125$
larger than in the mII simulation, corresponding to the same mass
resolution of the original Millennium simulation (\citealt{millenium}).\\
The initial conditions for the mmII are identical to those for the main (mII) run. This means that although they share the power spectrum of the fluctuation field, this is sampled with a factor of $125$ less particles in the case of the mmII; it is reasonable to expect that the smallest perturbations are then not represented. This reverses the situation with respect to the series of simulations analysed so far in this section. 
\begin{figure}
\includegraphics[width=84mm]{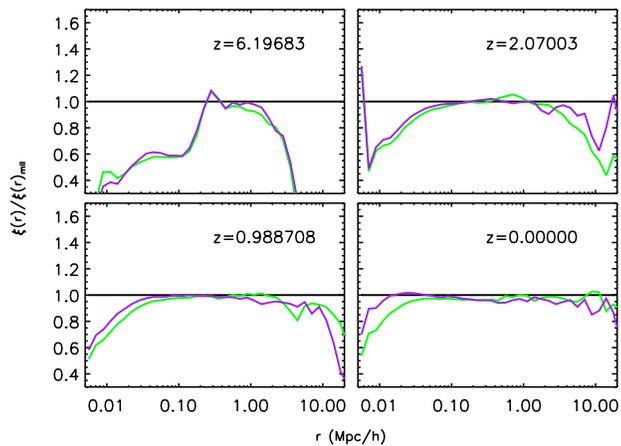}
 \caption{Comparison of the two-point correlation functions from the mII (black curve), mmII (green curve) and ``adaptive''-mmII (purple curve) simulations at different redshifts; shown is the ratio to the results from the mII. See the text for a description of the three cases.}
  \label{mmII_corr}
\end{figure}
Fig.~\ref{mmII_corr} shows the ratio of the correlation functions from the mmII and
``adaptive''-mmII to the original mII at four different redshifts. At high redshifts, down
to approximately $z=2$, the amplitude of $\xi(r)$ at scales smaller than $\simeq 20\; h^{-1}\mathrm{kpc}$ is lower in the
``adaptive''-mmII than in the mmII; the result is not surprising
though, as at those redshifts we expect almost all the particles to have softenings considerably larger than $\epsilon = 5\; h^{-1}\mathrm{kpc}$, the value adopted in the original mmII. At low redshifts the situation reverses, leading to a correlation functions which approaches more and more tightly the results
of the mII simulation; at $z=0$ and on scales below $\sim 10\; h^{-1}\mathrm{kpc}$, the ``adaptive'' correlation function is a factor $\sim 1.3$ above that of the mmII.
Fig.~\ref{mmII_mfct} shows the mass functions for the three simulations at the same redshifts as in Fig.~\ref{mmII_corr}. The halos are identified by the \fof algorithm, using $b=0.2$ and a minimum of $20$ particles. Again, the mass function are in agreement at all redshifts. \\
As for the tests presented before, this demonstrate that using adaptive gravitational softening while incorporating the correction term into the equation of motion allows to resolve the smallest scales better than simulations with fixed softening at the same resolution. The results converge to what is expected from much higher resolution simulations and no substantial degrading of the results is observed at high redshifts, where the formal resolution of the simulations using adaptive gravitational softening is naturally smaller than in the simulations with fixed softening.

\section{Conclusions}
\label{sec:concl}
We have implemented adaptive gravitational softening in the
cosmological TreePM code \gadgt. The formalism was introduced first by
\cite{PM07} and features the same technique used by SPH to determine
individual softening lengths for each of the simulation particles. The
spatial variation of this scale requires a modification of the
equation of motion governing the evolution of the particles'
trajectories in order to be consistent with the new dependencies
introduced in the system's Lagrangian.\\  We have applied this
technique to several test cases in order to check the behaviour of the
total energy when the new equation of motion is used; we then moved to
the cosmological scenario and, specifically, to the simulation of the
large-scale structure of the Universe, where the evaluation of the
effects of adaptive softening is complicated by the lack of an
analytical solution to the problem.\\ Our main conclusions are:
\begin{itemize}
\item The inclusion of the correction term in the equation of motion
  is essential to ensure the conservation of total energy.
\item In cosmological simulations of the large-scale structure a
  number of neighbours $\approx60$ is needed to obtain a converged
  estimation of the correction term.
\item With such a choice we show that -- in contrary to previous
  claims in literature -- the adoption of adaptive softening does not
  lead to an under-representation of halos at low masses, if the
  correct equation of motion is used.
\item Using the adaptive scheme effectively increases the dynamical
  range of cosmological simulations while the computational costs only
  mildly increase. Especially the amplitude of the two-point
  correlation function at small scales and the subhalo mass function
  improve compared to simulations with the same number of particles
  and fixed gravitational softening, anticipating the results obtained
  in higher-resolution simulations.
\item The convergence of the inner density profile for the most
  massive object found in the simulations improves significantly when
  the adaptive softening and the correct equation of motion is used.
\item When re-simulating a low-resolution version of the Millennium-II
  simulation \citep{mII} and comparing the results obtained with fixed
  and adaptive softening, we notice again perfect agreement in the
  mass functions at all times and an evolution of the ``adaptive''
  correlation function towards the higher amplitude of the
  Millennium-II's at late times.
\end{itemize}
We have so far applied the method to scenarios with equal-mass
particle species, either gas-only or dark-matter-only;  
simulations including both species with different masses would
also considerably benefit from the adoption of the scheme: the
adaptive behaviour of the resolution scale would allow to follow the
collapse of dark matter and particularly gas down to scales currently
unachievable in standard simulations at comparable resolution. 
However, having mixtures of particles with different masses is
a non-trivial extension of such scheme and will be investigated in future work.
Such future studies will be of particular importance considering that cold gas and star particles
in hydrodynamical simulations are likely to clump at lengthscales
below those typical chosen for the gravitational softening.
 
\section*{Acknowledgments}
We are grateful to Daniel J.~Price for precious clarifications on some aspects of the algorithm. We thank Michael Boylan-Kolchin for providing the initial conditions, mass functions and correlation functions for the results of Sec.~\ref{sec:mmII}. A special thank to Thorsten Naab for reading the manuscript and providing helpful suggestions. Finally, we are particularly grateful to Steffen Knollmann, whose comments have helped us improve the quality of this paper.
F.I. acknowledges useful discussions with Alessia Gualandris.
K.D. acknowledges the support by the DFG Priority Programme 1177 and additional support by the DFG Cluster of Excellence ``Origin and Structure of the Universe''.

\bibliography{biblio}{}
\bibliographystyle{mn2e}

\label{lastpage}

\end{document}